\documentclass[a4paper,fleqn,usenatbib]{mnras}

\usepackage{times}
\usepackage{graphicx}
\usepackage{hyperref}
\usepackage{amssymb}

\newcommand{\de}{{\rm d}}
\newcommand{\deriv}[2]{ \frac{{\rm d} #1}{{\rm d} #2} }
\newcommand{\pderiv}[2]{ \frac{\partial #1}{\partial #2} }

\defcitealias{fiacconi+16}{Paper~I}

\title[Internal rotation in quasi-stars]{
Light or heavy supermassive black hole seeds: the role of internal rotation in the fate of supermassive stars
}

\author[Fiacconi \& Rossi]{
Davide Fiacconi$^{1,}$\thanks{E-mail: fiacconi@physik.uzh.ch} 
and Elena M. Rossi$^{2,}$\thanks{E-mail: emr@strw.leidenuniv.nl}\\
$^{1}$Center for Theoretical Astrophysics and Cosmology, Institute for Computational Science, University of Zurich, Winterthurerstrasse 190,\\
CH-8057 Z\"{u}rich, Switzerland\\
$^{2}$Leiden Observatory, Leiden University, PO Box 9513, 2300 RA, Leiden, the Netherlands
}

\begin{document}

\label{firstpage}

\pagerange{\pageref{firstpage}--\pageref{lastpage}}

\maketitle


\begin{abstract}
Supermassive black holes are a key ingredient of galaxy evolution.
However, their origin is still highly debated.
In one of the leading formation scenarios, a black hole of 
$\sim100$~M$_{\sun}$ results from the collapse of the inner core of a supermassive star 
($\gtrsim 10^{4-5}$~M$_{\sun}$), created by the rapid accumulation ($\gtrsim 0.1 $~M$_{\sun}$~yr$^{-1}$) of 
pristine gas at the centre of newly formed galaxies at $z\sim 15$.
The subsequent evolution is still speculative: the remaining gas in the supermassive star can either 
directly plunge into the nascent black hole, or part of it can form a central accretion disc, 
whose luminosity sustains a surrounding, massive, and nearly hydrostatic envelope (a system called a ``quasi-star''). 
To address this point, we consider the effect of rotation on a quasi-star, as angular momentum is inevitably 
transported towards the galactic nucleus by the accumulating gas. 
Using a model for the internal redistribution of angular momentum that qualitative matches results from simulations of rotating convective stellar envelopes,
we show that quasi-stars with an envelope mass greater than a few $10^{5}$~M$_{\sun} \times 
(\rm black~hole~mass/100~M_{\sun})^{0.82}$ have highly sub-keplerian gas motion in their 
core, preventing gas circularisation outside the black hole's horizon. Less 
massive quasi-stars could form but last for only $\lesssim 10^4$ years before the accretion luminosity 
unbinds the envelope, suppressing the black hole growth. 
We speculate that this might eventually lead to a dual black hole seed population: (i) massive ($>10^{4}$~M$_{\sun}$) 
seeds formed in the most massive ($> 10^{8}$~M$_{\sun}$) and rare haloes; (ii) lighter ($\sim 10^{2}$~M$_{\sun}$) seeds 
to be found in less massive and therefore more common haloes.
\end{abstract}



\section{Introduction}

During the last ten years or so, observations have unambiguously proved the existence of supermassive black holes accreting at the centre of bright quasars at redshifts $z \gtrsim 6$ with masses in excess of $10^{9}$~M$_{\sun}$ \citep{fan+06,willott+10,mortlock+11, wu+15}.
Despite that those objects are not perhaps representative of the entire population of supermassive black holes at $z \gtrsim 6$ (e.g. \citealt{treister+13,weigel+15}), they represent a challenge for many theoretical models that attempt to describe the formation of the first black hole seeds.
Indeed, black hole seeds originating both as the leftovers of the first population (PopIII) stars (with masses $\lesssim 100$~M$_{\sun}$; \citealt{madau+01,tanaka+09}), and as the product of dynamical processes at the centre of primordial nuclear star cluster (with masses $\lesssim 1000$~M$_{\sun}$; \citealt{quinlan+90,devecchi+09,devecchi+12}), are not expected to grow fast enough to reach $\sim 10^{9}$~M$_{\sun}$ by $z\sim 6$ (e.g. \citealt{johnson+07,pelupessy+07,milosavljevic+09}), unless they experience prolonged periods of super-Eddington accretion (e.g. \citealt{madau+14,volonteri+15}).

A possible way out is to allow for the existence of massive black hole seeds ($\sim 10^4-10^5$~M$_{\sun}$) that can grow sub-Eddington and still match the masses of the quasars at $z\sim 6$.
This is achieved by the so called `direct collapse' scenario, according to which massive clouds ($\sim 10^{6} - 10^{7}$~M$_{\sun}$) of pristine gas can collapse almost isothermally at the centre of protogalactic, HI-cooling haloes (i.e. with virial temperature $T_{\rm vir} \gtrsim 10^4$~K; e.g. \citealt{bromm+03,begelman+06,lodato+06,latif+13,choi+13,choi+15}).
During the collapse, fragmentation can be avoided by dissociating H$_{2}$ (the main coolant in absence of metals) through the irradiation of Lyman-Werner photons coming from nearby, star-forming galaxies (e.g. \citealt{dijkstra+14,regan+14,agarwal+15}), while supersonic turbulence and non-axisymmetric perturbations can remove angular momentum from the collapsing gas and suppress fragmentation further \citep{begelman+09,choi+13,choi+15,mayer+15}.

However, even if the concurrency of all the processes above can be attained and it leads to the onset of the gravitational collapse, it is still unclear how the black hole seed would actually form.
The expectation is that the collapse proceeds almost isothermally at $\sim 8000$~K (as set by HI-line cooling) until a supermassive protostars forms at the fragmentation scale $\sim 10^5 - 10^6$~M$_{\sun}$, quickly accreting at $\sim 0.1-1$~M$_{\sun}$~yr$^{-1}$ \citep{hosokawa+12,hosokawa+13}.
After exhausting nuclear reactions, the central core of a supermassive star $\gtrsim 10^{4-5}$~M$_{\sun}$
is expected to collapse in a $\sim 100$~M$_{\sun}$ embryo black hole \citep{begelman+10,hosokawa+13} because of general relativistic radial instability \citep{baumgarte+99,shibata+02}.
The black hole is surrounded by most of the mass of the original envelope which is still contracting on a longer dynamical timescale.
It is unclear what happens next.
Possibly, the infalling gas retains enough angular momentum to build some kind of an accretion disc around the black hole at the centre of the envelope.
This structure can reach the equilibrium where the accretion luminosity is used to sustain the 
massive envelope against its own self-gravity, i.e. a \emph{quasi-star} 
(\citealt{begelman+08,ball+11,dotan+11}; \citealt[][hereafter Paper~I]{fiacconi+16}).
Therefore, a necessary ingredient for a quasi-star is the presence of a central accretion disc. 
It forms within the sphere of 
influence of the black hole ($\ll$ than the quasi-star radius) and it is able to convectively 
transport outward into the hydrostatic envelope the potential energy liberated through accretion.

In this way, quasi-stars can quickly grow their central black holes to $\sim 
10^{4}$~M$_{\sun}$ at (or above) 
the Eddington rate for the whole envelope, although strong outflows can limit the black hole growth 
(\citealt{dotan+11}; \citetalias{fiacconi+16}).
At the same time, the envelope keeps accreting mass from the environment.
Whether such accretion proceeds directly through filaments or from a protogalactic disc, the gas likely transports some amount of angular momentum that is transferred to the quasi-star and redistributed within it.
Quasi-stars are then expected to rotate, possibly faster on the equatorial plane than on the poles if they are embedded in a disc.

Rotation may have a few effects on the evolution of quasi-stars.
In analogy with normal stars, it could modify the internal structure of the quasi-star (e.g. 
\citealt{palacios+06, eggenberger+10,brott+11,ekstrom+12}), or it can stabilise the object against 
general relativistic instabilities, unless too massive ($\gtrsim 10^8$~M$_{\sun}$; 
\citealt{fowler+66}). Finally, a crucial feature that depends on the internal 
redistribution of angular momentum is the ability of the gas to circularise and to form an accretion disc. 
Here we explore whether the conditions for a disc to form are typically met in steady-rotating quasi-stars 
and we find that in most of the parameter space the answer is negative.
Although this result might be sensitive to environmental conditions as well as to details of the convective structures,
it opens in principle the possibility of directly forming massive seeds, without the intermediate stage of a quasi-star.

This paper is organised as follows.
In Section \ref{sec_2}, we present our analytical model to describe the differential rotation within 
quasi-stars and we calculate the angular velocity profiles, finding that the angular momentum at the 
boundary of the accretion region is typically much less than a percent of the Keplerian angular 
momentum at the same location.
Before concluding, we discuss in Section \ref{discussion} the speculative implications of our work, cautioning at the same time 
about the limitations of our approach.


\section{The model of rotating quasi-stars} \label{sec_2}


\subsection{Quasi-stars as loaded polytropes} \label{subsec_loaded_polytrope}

The hydrostatic structure of a quasi-star is constituted by a radiation-dominated, convective envelope, surrounded by a thin, radiative layer (\citealt{begelman+08,ball+11,dotan+11}; \citetalias{fiacconi+16}).
Since the envelope represents the majority of the mass and volume of a quasi-star and convective regions can be described accurately by an adiabatic temperature gradient, a quasi-star can be modelled as a polytropic gas with index $n=3$.
A polytropic gas is characterised by a barotropic equation of state $P(\rho) = P_{\rm c} (\rho / \rho_{\rm c})^{\gamma}$, where $P_{\rm c}$ and $\rho_{\rm c}$ are the central pressure and density, respectively, and the adiabatic index $\gamma = 1 + 1/n = 4/3$ for a radiation-dominated gas.
Polytropes are regular solutions of the Lane-Emden equations with inner boundary conditions in the standard dimensionless density and mass variables $\Theta_{\rm c} = \Theta(0) = 1$ and $\phi_{\rm c} = \phi(0) = 0$.
When $n<5$, they extend up to the dimensionless radius $\xi_{\star} = r_{\star}/\alpha$, where $\Theta_{\star} = 
\Theta(\xi_{\star}) = 0$ and $\alpha$ is the standard radial normalisation, and they enclose a total, finite mass $M_{\star} = 4 \pi \rho_{\rm c} 
\alpha^3 \phi_{\star}$ (e.g. \citealt{ball+12}).

Additionally, quasi-stars are characterised by the presence of a central black hole of mass 
$M_{\bullet}$.
We can model this feature by changing the inner boundary conditions: we assume that within the 
radius $r_{0}$, the enclosed mass is $M(r_{0}) = M_{\bullet}$ and that the density and the pressure 
are normalised to the values $\rho_{0}$ and $P_{0}$ at $r_{0}$, respectively.
The radius $r_{0}$ is the size of the gravitational sphere of influence of the black hole and
is typically of the order of its Bondi radius $r_{\rm B}$:
\begin{equation}\label{eq_r_0}
r_0 = b r_{\rm B} = b \frac{G M_{\bullet}}{2 c_{\rm s, 0}^2},
\end{equation}
where $c_{\rm s, 0}^2 = \gamma P_0 / \rho_0$ and $b$ is a numerical constant of the order of 
few.
In terms of dimensionless quantities, the new boundary conditions at 
$\xi_0 = r_0/\alpha$  are $\Theta(\xi_0) = \Theta_0 = 1$ and $\phi_0 = \phi(r_0) = M_{\bullet} / (4 
\pi \rho_0 \alpha^3)$.
A polytropic solution with non-zero central mass (i.e. with the latter boundary conditions) is called \emph{loaded polytrope} \citep{huntley+75}.
Throughout the rest of the paper, we use loaded polytropes to model the internal, hydrostatic structure of a quasi-stars assuming $n=3$.

\begin{figure}
\begin{center}
\includegraphics[width=\columnwidth]{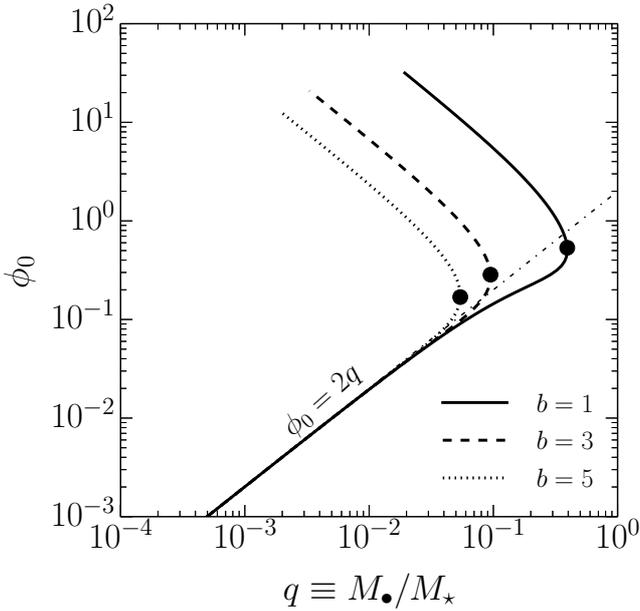}
\end{center}
\caption{Relation between the mass ratio $q = M_{\bullet}/M_{\star} = \phi_0 / \phi_{\star}$ and the dimensionless black hole mass $\phi_0$.
Continuous, dashed and dotted lines show the case $b=1$, $b=3$ and $b=5$, respectively.
The black dots show the position of the limiting mass ratio $q$.
The lower branch ($\phi_0 < 0.1$) is the only one that gives physical solutions (see the text) and it almost does not depend on $b$.
The dash-dotted line shows the relation $\phi_0 = 2 q$.
}
\label{fig_mass_ratio_phi0}
\end{figure}

We note that $\xi_0$ and $\phi_0$ are not independent, but they are related by:
\begin{equation}
\xi_0 = \frac{3b}{2} \phi_0.
\end{equation}
Therefore, the boundary conditions can be fully determined by choosing a value for $\phi_0$.
In turn, this is related through the Lane-Emden equation to the total mass of the envelope $\phi_{\star}$.
This relation is shown in Figure \ref{fig_mass_ratio_phi0} in terms of the the mass ratio $q \equiv M_{\bullet}/M_{\star} = \phi_0/\phi_{\star}$ as a function of $\phi_0$ for different values of $b$.
The mass ratio $q$ has always a maximum at $\phi_0 = \tilde{\phi}_0$.
This occurrence has been described in details by \citet{ball+12} as a generalisation of a 
Sch\"{o}nberg-Chandrasekhar-like limit for polytropic envelopes surrounding a central core 
\citep{schonberg+42}.
Quasi-stars have typically $q <10^{-2}$ \citepalias{fiacconi+16}. 
Solutions on the $\phi_0 
> \tilde{\phi}_0$ branch are unphysical because they 
reach zero mass before zero radius. Acceptable solutions lie on the $\phi_0 < \tilde{\phi}_0$ branch, 
where the dependency on $b$ becomes very weak. On this branch, 
we find empirically $\phi_0 \approx 2 q$, as shown in Figure \ref{fig_mass_ratio_phi0}.
From this relation, we can build any solution as follows.
First, we choose a value of $q$, typically between $\sim 10^{-4}$ and $\sim 10^{-2}$.
This maps to the value of $\phi_{0}$ necessary to set the boundary conditions and specify
$\phi_{\star}$.
We can then rescale the dimensionless solution with a specified $q$ to any solution in physical units by specifying the central black hole mass $M_{\bullet}$ and the pressure $P_{0}$.
The density $\rho_0$ can then be obtained as:
\begin{eqnarray}
\rho_0 & = & \displaystyle \left[ \frac{(n+1)^{3}}{4 \pi G^3} \right]^{1/4} \frac{\phi_0^{1/2} P_0^{3/4}}{M_{\bullet}^{1/2}} \approx \nonumber \\
& \approx & 1.2 \times 10^{-5}~q^{1/2}~p_{0,8}^{3/4}~m_{\bullet}^{-1/2}~{\rm g~cm^{-3}},
\end{eqnarray} 
where we use $n=3$, $M_{\bullet} = m_{\bullet}$~M$_{\sun}$, and $\phi_0 \approx 2 q$.
We discuss the limitations of this simplified treatment of the interior of quasi-stars in Section \ref{subsec_limitations}.


\subsection{Differential rotation inside quasi-stars} \label{subsec_internal_rotation}

In a recent series of papers, Balbus and collaborators have developed a theory to describe the
convective zone in the Sun \citep{balbus+09,balbus+10,balbus+10b,balbus+12,balbus+12b}.
Their model successfully reproduces the isorotation contours within the solar convective zone and 
the tachocline from the helioseismology data of the Global Oscillation Network Group (GONG).
Here, we review the main features of the model and we then apply it to quasi-stars, 
mostly following \citet{balbus+09} and \citet{balbus+10}.
We also verify in Section \ref{subsec_limitations} the applicability of this model to the quasi-star case.

We consider an azimuthally rotating, convective gas flow (generically a star) in spherical coordinates $(r, \theta, \phi)$, where $r$ is the radial distance from the centre, $\theta$ is the colatitude angle and $\phi$ is the azimuthal angle.
The flow is symmetric with respect to the rotation axis, i.e. the thermodynamic variables characterising it, such as the density $\rho$, the pressure $P$ and the specific entropy $s$, do not depend on $\phi$, but they generally depend on $r$ and $\theta$.
The only velocity component is the azimuthal velocity $v_{\phi} = r~\sin \theta~\Omega(r,\theta)$, where $\Omega(r,\theta)$ is the angular velocity.
We neglect any departure from sphericity, implicitly assuming slow rotation.
Such a flow in steady state is described by the following Euleur equations 
($r$- and $\theta$-component, respectively, while the azimuthal component is $0=0$):
\begin{equation}
\left\{\begin{array}{rll}
\displaystyle -\frac{1}{\rho} \pderiv{P}{r}-  \pderiv{\Phi}{r} & = & 0,\\
\displaystyle \frac{v_{\phi}^2 \cot{\theta}}{r}-\frac{1}{\rho} \pderiv{P}{\theta}-\frac{1}{r} 
\pderiv{\Phi}{\theta} & = & 0,
\end{array}\right.
\label{eq:euleur}
\end{equation}
where $\Phi$ is the gravitational potential.
Note that in the radial direction we neglect the (weak) centrifugal force\footnote{In fact, 
this approximation is only necessary to derive equation (\ref{eq_TWE}) {\it after} the $\phi$-component 
of the curl of those Euler equations has been taken.
For clarity, however, we already drop the centrifugal force at this early step.}.
By taking the $\phi$-component of the curl of the Euler equations (equation \ref{eq:euleur}), and 
dropping terms 
proportional to $\partial P /\partial \theta \ll \partial P / \partial r$, we obtain the thermal 
wind equation \citep{kitchatinov+95,thompson+03,balbus+09, balbus+10, balbus+12}:
\begin{equation}\label{eq_TWE}
\pderiv{\Omega^2}{r} - \frac{\tan \theta}{r} \pderiv{\Omega^2}{\theta} = 
\frac{1}{\gamma~r^2~\sin\theta~\cos\theta} \deriv{\Phi}{r} \pderiv{\sigma}{\theta},
\end{equation}
where we have introduced the dimensionless entropy function:
\begin{equation}
\sigma = \log \left[ \frac{P}{P_0} \left( \frac{\rho}{\rho_0} \right)^{-\gamma} \right],
\end{equation}
which is proportional to (or monotonically dependent on) $s$.
Equation (\ref{eq_TWE}) neglects the contribution from convective turbulence to the 
velocity field (it is in fact a time-averaged description of the flow) and it is not valid for 
highly 
magnetised stars, but a weak magnetic field can be accommodated \citep{balbus_solo+09}.

Let us now introduce the {\em residual} entropy: the azimuthally averaged entropy profile left, 
after the radial profile $\sigma_{r}$ has been subtracted off:
\begin{equation}
\sigma'(r,\theta) = \sigma - \sigma_{r}(r).
\end{equation}
Since equation (\ref{eq_TWE}) depends on $\sigma$ exclusively through its $\theta$ derivative, 
the differential profile $\Omega(r,\theta)$ could be determined after knowing $\sigma'$, regardless of $\sigma_r$.
Convection in a non-rotating star establishes a stable 
entropy radial profile $\sigma_r$, as a result of an equilibrium reached between central stellar heating and 
heat transport. Convective cells moves on average along the radial direction. If now 
a small amount of rotation is added, the convective cells will tend, on average, to drift towards 
surfaces of constant angular rotation. This is because differential rotation tends to confine 
the flow in sheet of constant $\Omega$. This assumes of course that the rotational surfaces can 
effectively interact with the convective cells during their lifetime, which is reasonable if 
they are long lasting structures. In the presence of a relative small degree of rotation, we can 
therefore argue that $\sigma_{r}$ is similar to that established in a non-rotating star, while 
$\sigma'(r,\theta)$ is a small departure from $\sigma_{r}$, closely connected to the differential 
rotational profile within the star. 
Following \citet{balbus+09}, we assume, $\sigma' = f(\Omega^2)$ 
which implies  
that surfaces of constant residual entropy coincide with surfaces of constant angular 
velocity. Though still not unambiguously demonstrated, this conjecture provides remarkable results 
when used to describe the solar convective zone \citep{balbus+10b,balbus+12,balbus+12b}.
In addition, it is also supported, at least qualitatively, by the results of hydrodynamical 
simulations showing similarity between constant $\Omega$ and $\sigma'$ contours 
(\citealt{miesch+06}; see also figure 2 from \citealt{balbus+09}).

With this relation, $\sigma' = f(\Omega^2)$, equation (\ref{eq_TWE}) can be rewritten as:
\begin{equation} \label{eq_TWE_balbus}
\pderiv{\Omega^2}{r} - \left( \frac{\tan \theta}{r} + \frac{f'}{\gamma~r^2~\sin\theta~\cos\theta} \deriv{\Phi}{r}  \right) \pderiv{\Omega^2}{\theta} = 0,
\end{equation}
where $f' = \de \sigma' / \de \Omega^2$.
The above equation has the form $\bmath{u} \cdot \nabla \Omega^2 = 0$, where $\bmath{u}$ is the
vector tangential to the surfaces of constant $\Omega$ (i.e. it is their ``velocity'' vector).
Such surfaces $\bzeta = (r, \theta(r))$ can be obtained by integrating the ordinary differential 
equation $\dot{\bzeta} 
= \bmath{u}$ (where $\dot{~}$ indicates the derivative with respect to any dummy parameter). More 
practically, one divides the polar and radial component of that vectorial equation and obtains 
the following single equation:
\begin{equation} \label{eq_surface}
\deriv{\theta(r)}{r} =   - \frac{\tan \theta}{r} - \frac{f'}{\gamma~r^2~\sin\theta~\cos\theta} 
\deriv{\Phi(r)}{r}.
\end{equation}
If we recall that $f'$ depends only on $\Omega^2$ and that equation (\ref{eq_surface}) describes 
surfaces of constant $\Omega^2$, we can finally integrate the above equation considering $f'$ as 
constant:
\begin{equation}
r^2 \sin^2\theta = A - \frac{2 f'}{\gamma} \Phi(r),
\end{equation}
where $A$ is an integration constant.
These iso-$\Omega^2$ surfaces are the characteristics of equation (\ref{eq_TWE_balbus}).
Note that on each surface, $f'$ can assume a different constant value.

We can determine the constant $A$ by specifying a starting position for each characteristic.
We take the position $(R_{\star}, \theta_{\star})$ at the surface of a spherical star with radius $R_{\star}$ and we obtain:
\begin{equation}\label{eq_characteristic}
r^2 \sin^2\theta = R_{\star}^2 \sin^2 \theta_{\star} - \frac{2 f'}{\gamma} (\Phi(r) - \Phi(R_{\star}) ),
\end{equation}
where now $f' = f'(\Omega(R_{\star}, \theta_{\star}))$ has to be specified and the internal structure of the star influences the result through $\Phi$.
The curves described by equation (\ref{eq_characteristic}) are constant $\Omega$ contours, 
therefore they can be used to reconstruct the 2-dimensional $\Omega(r,\theta)$ by assigning a value of 
$\Omega$ at a given radius. Specifically, we will supply $\Omega_{\star}(\theta_{\star})$ at 
$R_{\star}$. We can then isolate $\theta_{\star}(r,\theta)$ from equation (\ref{eq_characteristic}) 
and obtain $\Omega(r,\theta) = \Omega_{\star}(\theta_{\star}(r,\theta))$.

We can now use this method to explicitly calculate the internal differential rotation of 
quasi-stars,  once we specify their internal structure.
Since we describe quasi-stars as loaded polytropes (see Section \ref{subsec_loaded_polytrope}), we
can integrate the equation of hydrostatic equilibrium between $r$ and $R_{\star}$ for a polytropic equation of state and obtain:
\begin{equation}
\Phi(r) - \Phi(R_{\star}) = - \frac{c_{\rm s}^2}{\gamma - 1} = - 3~c_{\rm s, 0}^2~\Theta_{3}(r; q),
\label{eq:integrated_polytrope}
\end{equation}
where $c_{\rm s,0}^2 = \gamma P_0 / \rho_0$ is the sound speed at $r_0$, $\Theta_3(r; q)$ is the 
loaded polytrope solution for $n=3$ and a given $q$, and $\gamma = 4/3$.
Substituting equation (\ref{eq:integrated_polytrope}) into equation (\ref{eq_characteristic}), we finally get:
\begin{equation}
\sin^2\theta_{\star} = \left(\frac{r}{R_{\star}} \right)^2 \sin^2 \theta -\beta~\Theta_3(r; q),
\label{eq:diff_rotation_qs}
\end{equation}
where we define:
\begin{equation} \label{eq_beta}
\beta = \frac{9 c_{\rm s,0}^2 f'}{2 R_{\star}^2}.
\end{equation}

There is still a quantity that has remained general in our treatment, namely $f'(\Omega^2)$.
Unfortunately, we do not know a priori its functional form, and only hydrodynamical simulations of global 3D convection could clarify this point.
However, to avoid unnecessary mathematical complication at this stage, we assume the simplest functional form, i.e. a global constant for $f'$.
This simple choice is also motivated by the lack of any observational constraint; yet, such a choice is quite effective for the case of the Sun \citep{balbus+10b,balbus+12}.
However, we need to use additional reasonable arguments to constrain in our case the constant parameter $\beta$, that directly depends on $f'$ through equation (\ref{eq_beta}).

First, we expect $\beta <0$ (i.e. $f'<0$, like in the convective envelope of the Sun), 
since this implies slower rotating poles with respect to the equatorial regions. 
This configuration may naturally comes about when quasi-stars are fed by protogalactic discs near the equator, i.e. angular momentum is injected by the infalling material near the equator and has to be redistributed from there to the poles. 
Finally, we can also estimate the value of $|\beta|$ by recalling that $\sigma'$ is a small perturbation on 
the otherwise spherically symmetric entropy profile $\sigma_r$ which arises when the star rotates:
\begin{equation}
\sigma' \sim \frac{T_{\rm rot}}{U} \sim \frac{R_{\star}^2 \Omega^2 }{c_{\rm s,0}^2},
\end{equation}
where $T_{\rm rot} \sim M_{\star} R_{\star}^2 \Omega^2$ and $U \sim M_{\star} c_{\rm s,0}^2$ are the rotational kinetic energy and the gaseous internal energy of the star, respectively.
Therefore, $f' \sim \sigma'/\Omega^2 \sim R_{\star}^2 / c_{\rm s, 0}^2$ implies that $|\beta| \sim 
9/2 \sim$ a few.
Although this simple line of reasoning does not prove that $f'$ should be constant, it provides a gross estimate of the value of $|\beta|$ \emph{if} $f'$ is assumed to be constant.
However, we show in Section \ref{subsec_results} that the exact value of $|\beta|$ has a weak impact on our final conclusions and we discuss the limitations of our approach in Section \ref{subsec_limitations}.

\begin{figure*}
\begin{center}
\includegraphics[width=2\columnwidth]{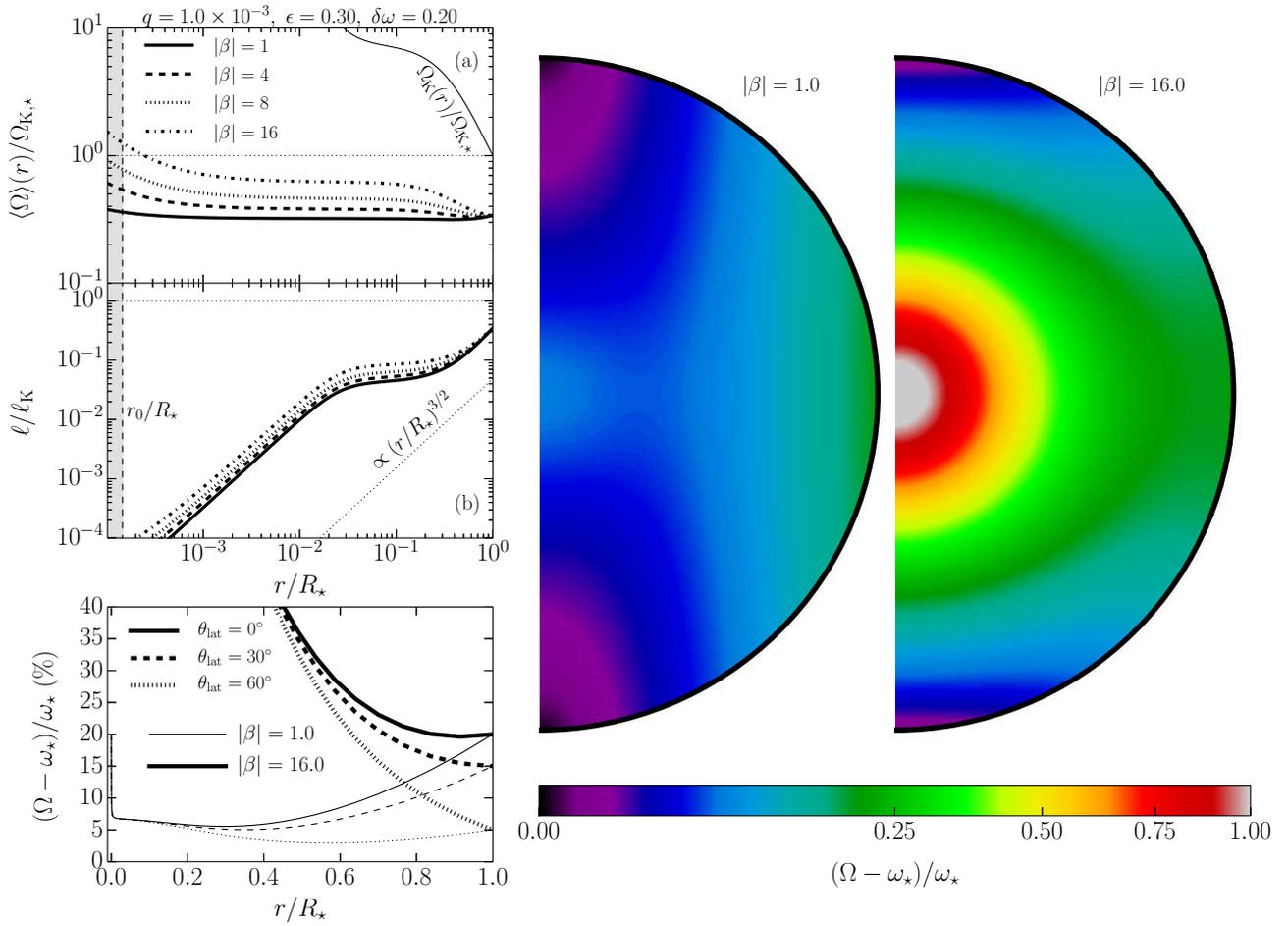}
\end{center}
\caption{Angular velocity for a quasi-star model characterised by $q=10^{-3}$, $\epsilon=0.3$ and $\delta \omega = 0.2$.
Top-left panel: radial profile of the $\theta$-averaged angular velocity (sub-panel (a)), normalised to the Keplerian angular velocity $\Omega_{\star}$, and of the ratio $\ell/\ell_{\rm K}$ of the $\theta$-averaged specific angular momentum over the Keplerian one for the same mass distribution (sub-panel (b)).
Solid, dashed, dotted, and dash-dotted lines correspond to $|\beta|=1$, 4, 8, 16, respectively.
The solid thin line in sub-panel (a) shows the Keplerian angular velocity.
Bottom-left panel: percentage excess of rotation relative to the polar angular velocity as a function of radius at constant latitude $\theta_{\rm lat}$.
Solid, dashed, and dotted lines correspond to $\theta_{\rm lat} = 0^{\circ}$ (equator), $30^{\circ}$, and $60^{\circ}$, respectively, while thin and thick lines refer to $|\beta|=1$ and $|\beta|=16$.
Right panel: 2D maps of the internal relative excess of angular velocity relative to the polar rotation for $|\beta|=1$ and $|\beta|=16$.
Note that the surface rotation of both maps is the same.
}
\label{fig_ang_mom}
\end{figure*}


\subsection{Angular velocity structure of quasi-stars} \label{subsec_results}

To explicitly calculate the differential rotation within a quasi-star, we need to specify the boundary conditions of the problem, i.e. 
the differential rotation at the surface $\Omega_{\star}(\theta_{\star})$.
We use a simple parametrisation of the form:
\begin{equation} \label{eq_ang_vel_surf}
\Omega_{\star}(\theta_{\star}) = \omega_{\star} (1 + \delta \omega \sin^2 \theta_{\star}),
\end{equation}
where $\omega_{\star}$ is the polar rotation, limited by the Keplerian velocity of the star $\Omega_{\rm K, \star}$, and $\delta \omega$ is the relative, fractional excess of rotation at the equator, with the limit $0 < \delta \omega < \epsilon^{-1} -1$, where $\epsilon = \omega_{\star} / \Omega_{\rm K, \star} < 1$.
This parametrisation of the differential rotation has been used to describe the Sun as well as other 
stars, with typical values $\delta \omega \sim 0.1$  (e.g. \citealt{balbus+09, reinhold+13}).

Figure \ref{fig_ang_mom} shows the angular velocity profiles and maps for a reference quasi-star with $q = 10^{-3}$ 
(e.g. a massive quasi-star with $M_{\bullet} = 10^3$~M$_{\sun}$ and $M_{\star} = 10^{6}$~M$_{\sun}$), 
$\epsilon = 0.3$ and $\delta\omega = 0.2$ (i.e. rotating at $0.36~\Omega_{\rm K, \star}$ at the equator, with a differential velocity of 20\%
between the equator and the poles), and we vary the value of $|\beta|$ between 1 and 16.
The upper-left panel (sub-panel (a)) shows the radial profile of the $\theta$-averaged angular velocity $\langle \Omega \rangle$ (normalised by $\Omega_{\rm K, \star}$), highlighting its behaviour at small radii.
Initially, $\langle \Omega \rangle$ grows from the surface of the star inward for most of the stellar volume till $\sim 0.1 R_{\star}$.
This growth is accentuated for larger values of $|\beta|$.
Within $0.1 R_{\star}$, $\langle \Omega \rangle$ remains almost constant, assuming a solid-body-like rotation law and 
following the central density of the gas that also starts to flatten in a central core. 
However, the angular velocity deviates from constant within $0.005~R_{\star}$ (or $\sim 20~r_{0}$) because of the 
presence of the central black hole, steepening at smaller radii.
The typical trend is $\propto r^{-\zeta}$, with $\zeta \sim 0.2 - 0.6$, increasing with $|\beta|$.
We compare $\langle \Omega \rangle$ with the Keplerian angular velocity $\Omega_{\rm K}(r) = \sqrt{ G M(r)/r^3}$ associated to 
the same mass distribution.
Outside $\sim 0.2~R_{\star}$, $\Omega_{\rm K}$ grows inward as $r^{-3/2}$ (since most of the mass of the envelope is contained in the central core), faster than $\langle \Omega \rangle$.
Then, it slightly flattens, but it suddenly starts to grow again as $r^{-3/2}$ due to the presence of the central black hole that dominates
the enclosed mass $\phi$ out to $\sim 0.02 R_{\star}$, resulting in $\Omega_{\rm K} \gg \langle \Omega \rangle$ at $r_0$.

Although convection can induce solid body rotation, this is not achieved in the entire envelope, but only in the central part.
This is shown in the lower-left panel of Figure \ref{fig_ang_mom}, where we plot the radial profiles of $\Omega$ (shown as 
the percentage excess of rotation compared to the surface angular velocity at the poles $\omega_{\star}$) at different latitudes $\theta_{\rm lat}$ ($\theta_{\rm lat} = 0^{\circ}$ means the equator).
Most of the stellar volume is differentially rotating  at different latitudes, as shown by the two extreme examples $|\beta|=1$ and $|\beta|=16$.
Those are representative of the two limiting cases: when $|\beta| \rightarrow 0$, the angular velocity becomes constant on cylinders.
This can be seen in the region close to the surface around the equator of the map corresponding to $|\beta|=1$.
At constant latitude, $\Omega$ decreases as $r$ goes from the surface to $\sim 0.3-0.4R_{\star}$, when it starts to mildly grow inward and it becomes nearly constant within $\sim 0.1 R_{\star}$; then, it steepens again close to the central black hole.
On the other hand, when $|\beta| \gg 1$, the angular velocity tends to be ``shellular'', i.e. it mostly follows the isobars and $\Omega$ varies with $r$ only.
That can be seen in the example map for $|\beta|=16$ within $\sim 0.5 R_{\star}$, while in the outermost parts of the star the angular velocity maintains a net $\theta$-dependency and it quickly grows as $r$ decreases.

The ratio $\ell / \ell_{\rm K}$ between the specific angular momentum $\ell$ and the Keplerian angular momentum 
for the same mass distribution closely relates to $\langle \Omega \rangle$ and $\Omega_{\rm K}$, as shown in the top-left panel (sub-panel (b)).
Outside $\sim 0.2 R_{\star}$, $\ell / \ell_{\rm K}$ first decreases inward, then flattens till $\sim 0.02 R_{\star}$, to decrease again roughly as $\propto r^{3/2}$.
Finally, within $0.005 R_{\star}$, the ratio decreases inward more slowly, as a consequence of 
the steepening of $\langle \Omega \rangle$ close to the central black hole (Figure \ref{fig_ang_mom}).
Close to $r_{0}$, $\ell/\ell_{\rm K}$ typically assumes values $\sim 10^{-4}$, with mild variations within a factor $\gtrsim 2$ when $|\beta|$ is changed between 1 and 16, suggesting that the exact value of $|\beta|$ has a minor impact on the ratio
 $\ell/\ell_{\rm K}$ at about $r_0$.
 
\begin{figure}
\begin{center}
\includegraphics[width=\columnwidth]{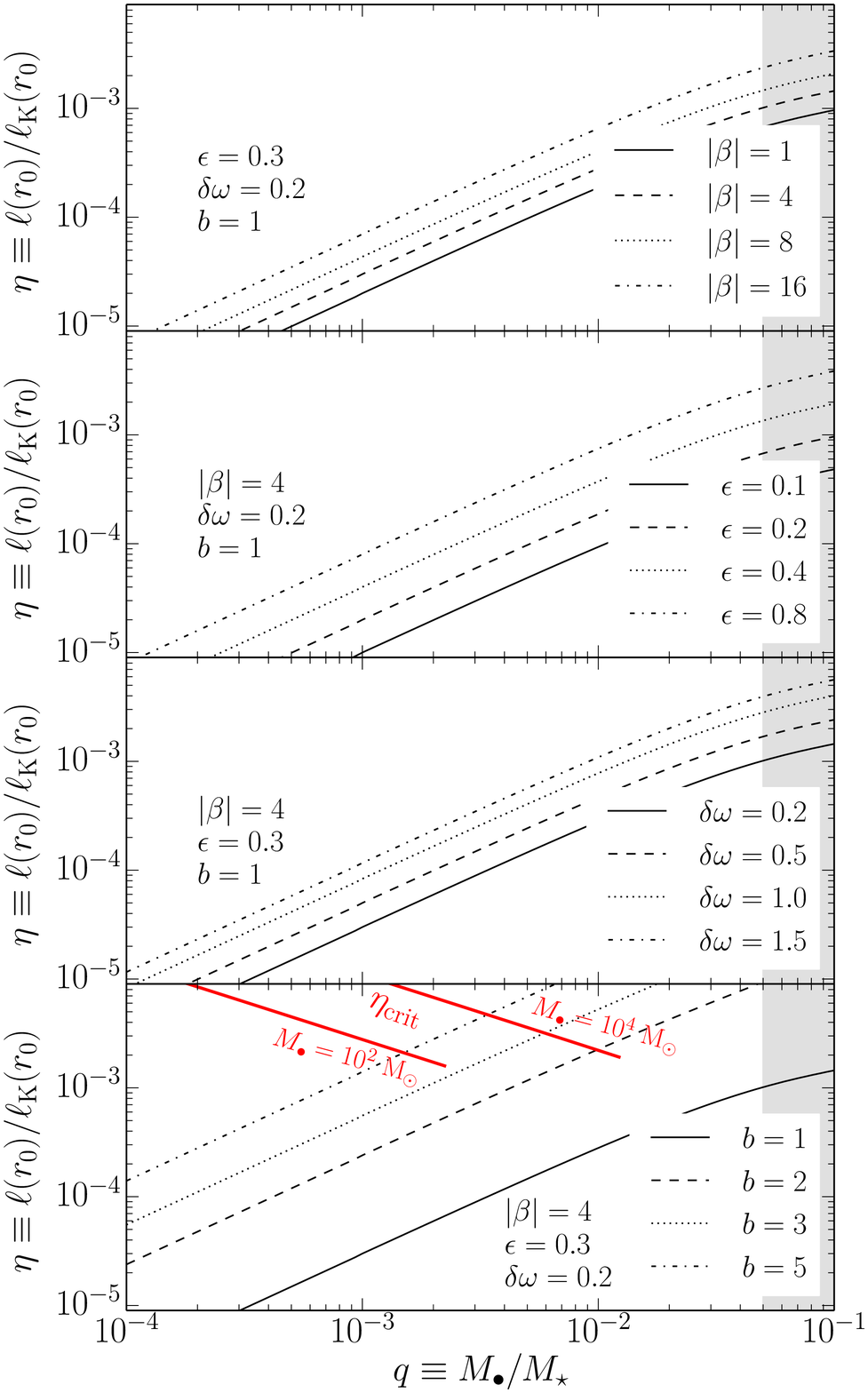}
\end{center}
\caption{Ratio $\ell / \ell_{\rm K}$ of the specific angular momentum over the Keplerian one at $r_0$ as a function of $q$ for different parameters.
From top to bottom: the effect of changing $\beta$, $\epsilon$, $\delta\omega$, and $b$.
The lower panel also shows two example curves (red continuous lines) for $\eta_{\rm crit}$ 
(see equation \ref{eq_ang_mom_constrain}), calculated from the models of \citetalias{fiacconi+16} for $M_{\bullet} = 10^{2}$ and $10^4$~M$_{\sun}$.
}
\label{fig_parameters}
\end{figure}
 
 The small $\ell/\ell_{\rm K}$ value at $r_0$ is a direct consequence of the quasi-star structure, and in particular of the presence of the central black hole.
This can be better understood by comparing a quasi-star with a similar system, namely a standard $\gamma = 4/3$ polytrope with the same envelope mass, but without the central black hole.
The pure polytropic structure is more compact and has a solid body rotation all the way to the centre, 
with a lower and nearly constant Keplerian velocity due to the absence of the black hole.
This case leads to a typical ratio $\ell / \ell_{\rm K}\sim 0.1-0.01$ larger then in the equivalent quasi-star.

We have tested the sensitivity of $\ell / \ell_{\rm K}$ around $r_0$ on the structural parameters of the star: $q$, $\epsilon$, $\delta \omega$, and $b$.
Figure \ref{fig_parameters} shows that quasi-stars with proportionally larger black holes at the centre (i.e. with larger $q$) 
tend to have 
larger $\ell/\ell_{\rm K}$ close to $r_0$, though this ratio remains confined within $\sim 10^{-2}$ for $q \leq 10^{-2}$ for all the parameter combinations that we expect to bracket consistent quasi-star solutions.
$\ell/\ell_{\rm K}$ departs more from $\propto r^{3/2}$ for larger $q$, becoming shallower close to $r_{0}$.
When the quasi-star rotates proportionally faster at the equator than at the pole (i.e. when 
$\epsilon$ decreases and $\delta \omega$ increases) the spread between the rotation laws with different $|\beta|$ 
increases, but this does not change their shape and the typical $\ell/\ell_{\rm K}$ close to $r_0$.
We note that the value of $\ell/\ell_{\rm K}$ at $r_0$ is mostly sensitive to $b$ (i.e. the size of the accretion region around the black hole, see equation \ref{eq_r_0}).
The ratio $\ell/\ell_{\rm K}$ at $r_0$ grows with $b$ as $r_0$ proportionally increases.
Nonetheless, $\ell/\ell_{\rm K}$ always remains well below 1, reaching up to $\sim 10^{-2}$ for $b=5$ for quasi-stars with large mass ratios $q \gtrsim 10^{-2}$.

We therefore conclude that, regardless of the strict values of the 
parameters assumed, the typical specific angular momentum where the gravity of the black 
hole starts to dominate (i.e. around $r_{0}$) is much lower than the local Keplerian angular momentum.
We discuss the implications (as well as the limitations) of this result for quasi-stars in Section~\ref{discussion}.


\section{Discussion and conclusions} \label{discussion}


\subsection{Possible implications}\label{subsec_implications}

\begin{figure}
\begin{center}
\includegraphics[width=\columnwidth]{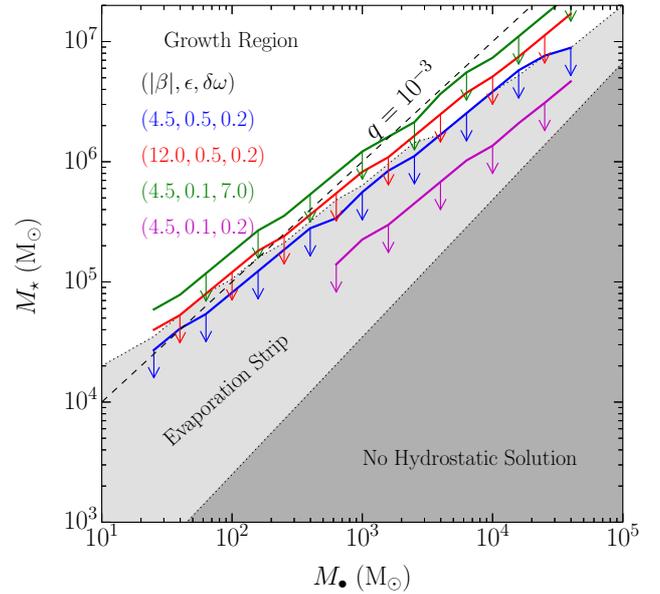}
\end{center}
\caption{$M_{\bullet} - M_{\star}$ plane for quasi-stars, divided in three regions: the growth region (white), the evaporation strip (light grey), and the no-hydrostatic solution region (dark grey).
Blue, red, green and magenta continuous lines show the upper limits on $M_{\star}$ as function of $M_{\bullet}$ for different choices of parameters, namely $(|\beta|, \epsilon, \delta \omega) = (4.5, 0.5, 0.2)$, $(12, 0.5, 0.2)$, $(4.5, 0.1, 7)$, and $(4.5, 0.1, 0.2)$, respectively.
Above this limits, which are computed from the models of \citetalias{fiacconi+16}, the condition $\eta > \eta_{\rm crit}$ is not satisfied.
For reference, the black dashed line corresponds to $q=10^{-3}$.
}
\label{fig_mbh_mstar}
\end{figure}

In this paper, we investigate the role of rotation within quasi-stars.
Although a fully self-consistent description of the rotating envelope is beyond the purposes of 
this paper, our treatment of rotation can be used to assess the coupling between the inner accretion 
region and the massive envelope.
Before discussing the implications of our findings for black holes forming inside rotating 
quasi-stars, we recall their fate, when rotation is not included. This is summarised in 
Figure \ref{fig_mbh_mstar}, adapted from \citetalias{fiacconi+16}, to which we refer the reader for more details. 
Combinations of $(M_{\bullet},M_{\star})$ that lie in the region marked as ``No Hydrostatic Solutions'' 
cannot form a stable envelope surrounding the black hole and therefore this latter cannot go through 
a phase of super-Eddington growth inside a quasi-star. For this to happen, the envelope needs to be 
at least a few hundred times more massive than its black hole (the parameter space marked as 
``Growth Region''). There, black holes with an initial mass of $\sim 100$~M$_{\sun}$ can reach in 
just $\gtrsim 10^{4}$ yr more than $10^4$~M$_{\sun}$, depending on the initial envelope mass. This is 
because the black hole accretes at or beyond the Eddington rate for {\it the envelope mass}. 
Moreover, in these massive envelopes the loss of mass via winds induced by the super-Eddington luminosities 
proceeds at a lower rate than the black hole growth.
The opposite is true for lower envelope masses at the same black hole masses within the 
``Evaporation Strip'', 
where outflows remove matter from the envelope faster than black hole accretion, and the latter 
is then suppressed. Here a quasi-star can form, but it can just last 
for $< 10^{4}$~yr, with little impact on the embedded black hole. 
We can now turn our attention to a discussion on how our results might affect this picture.

In our calculations, we neglect any general relativistic effect.
It is then worth comparing the Schwarzschild radius $r_{\rm s}$ of the central black hole,
that sets the size of the black hole horizon,
with the envelope's inner radius $r_0$, where our calculation stops:
\begin{eqnarray}\label{eq_r_schwarzschild}
\frac{r_{\rm s}}{r_{0}} & = & \displaystyle \frac{4 c_{\rm s,0}^2}{b c^2} = \frac{8 \left( \pi G^3 \right)^{1/4}}{3~c^{2}}~\frac{M_{\bullet}^{1/2}~P_0^{1/4}}{b~\phi_0^{1/2}} \approx \nonumber \\
& \approx & 5.2 \times 10^{-8}~b^{-1}~q^{-1/2}~m_{\bullet}^{1/2}~p_{0,8}^{1/4},
\end{eqnarray}
where we used $\phi_0 \approx 2 q$. Inserting consistent mass-pressure values from the
models shown in Figure \ref{fig_mbh_mstar} 
we typically find that $r_0$ is between 
a few to several thousand Schwarzschild radii of the black hole. For example, 
when we consider the Growth Region and take (i) a relatively small quasi-star, 
($M_{\bullet} = 100$~M$_{\sun}$, $M_{\star} = 2\times 10^{5}$~M$_{\sun}$ and $P_0 = 
6.3 \times 10^{10}$~erg~cm$^{-3}$); and (ii) a massive quasi-star with a 
relatively more massive black hole, ($M_{\bullet} = 10^{4}$~M$_{\sun}$, 
$M_{\star} = 10^{7}$~M$_{\sun}$ and $P_0 = 3.4 \times 
10^{10}$~erg~cm$^{-3}$), 
we find $r_{\rm s}/r_0 \approx 1.2 \times 10^{-4}$ and $r_{\rm s}/r_0 \approx 7.1 \times 10^{-4}$, 
respectively.
These estimates support our 
choice of neglecting any general relativistic effect and we can therefore safely use our 
results at $r_0$ to put boundary conditions to the central accretion flow.

Although a detailed modelling of the central accretion flow is
beyond the purpose of this work, we can still gain insight into its formation and some 
possible features from simple inferences from our results.
The results of Section \ref{subsec_results} suggest that the specific angular momentum at $r_{0}$ is $\ell_{0} = \eta~\ell_{\rm K}(r_{0})$, where $\eta$ is $\sim 10^{-3} - 10^{-4}$.
By assuming the conservation of angular momentum, we can calculate the circularisation radius 
$r_{\rm circ}$ around the central black hole, i.e. the radius at which $\ell_{0}$ corresponds to
a circular orbit:
\begin{equation}\label{eq_r_circ}
\frac{r_{\rm circ}}{r_{0}} = \eta^2,
\end{equation}
where we assume that $\Omega \simeq  
\Omega_{\rm K}$ below $r_{0}$, as the black hole's gravity dominates.
This radius tells us the scale below which some sort of accretion disc may eventually form. 
That requires $r_{\rm circ} > r_{\rm isco} \approx r_{\rm 
s}$, where $r_{\rm isco}$ is the radius of the innermost stable circular orbit, which is a few times 
$r_{\rm s}$, depending on the black hole spin.
We can combine equation (\ref{eq_r_schwarzschild}) and (\ref{eq_r_circ}) to determine a condition on 
$\eta$,
\begin{eqnarray}\label{eq_ang_mom_constrain}
\eta & > &\eta_{\rm crit} =  \displaystyle \left[ \frac{8 \left( \pi G^3 \right)^{1/4}}{3~c^{2}} \right]^{1/2} ~\frac{M_{\bullet}^{1/4} P_0^{1/8}}{b^{1/2} \phi_0^{1/4}} \approx \nonumber \\
& \approx & 1.3 \times 10^{-2}~b^{-1/2}~q_{-4}^{-1/4}~m_{\bullet, 100}^{1/4}~p_{0,10}^{1/8},
\end{eqnarray}
where $q = q_{-4} \times 10^{-4}$, $M_{\bullet} = m_{\bullet, 100} \times 100$~M$_{\sun}$, and $P_0 = p_{0,10} \times 10^{10}$~erg~s$^{-1}$.
When $\eta > \eta_{\rm crit}$, the gas circularisation is such that $r_{\rm circ} > r_{\rm isco}$ 
and an accretion disc can form at the centre of a quasi-star.
As an example, we calculate $\eta_{\rm crit}$ for the same quasi-star models used above, 
and find that $\eta_{\rm crit} \approx 1.1 \times 10^{-2}$ and $\eta_{\rm crit} \approx 2.7 
\times 10^{-2}$, respectively.
Those numbers are also representative of the whole Growth Region, as they weakly 
depend on $q$, $M_{\bullet}$ and $P_0$ (see equation \ref{eq_ang_mom_constrain}).
Interestingly, $\eta_{\rm crit} \gtrsim 10^{-2}$ is comfortably larger than the indicative upper limit 
on $\eta \sim \mbox{a few} \times 10^{-3}$ that we estimate in Section \ref{subsec_results}, 
leading to the conclusion that typical quasi-stars in the Growth Region 
might \emph{not} be able to develop an accretion disc at their centre. 

To better assess this point, we exploit the models used to construct Figure \ref{fig_mbh_mstar} 
to thoroughly explore the parameter space.
Specifically, we take the values of $M_{\bullet}$, $M_{\star}$ (hence $q$), and $P_0$ (calculated at
$r_0 = 5 r_{\rm B}$) and we use them to calculate $\eta$ and $\eta_{\rm crit}$ 
across the $M_{\bullet}-M_{\star}$ plane.
As discussed in Section \ref{subsec_results}, the value of $\eta$ depends on some parameters, namely 
$(|\beta|, \epsilon, \delta\omega)$.
We tested several configurations: (i) a ``fiducial'' model with $(4.5, 0.5, 0.2)$, (ii) a model with a higher value for $|\beta|$, $(12, 0.5, 0.2)$, (iii) a rapidly and differentially rotating quasi-star with $(4.5, 0.1, 7.0)$, and (iv) a slowly rotating quasi-star with $(4.5, 0.1, 0.2)$.
In all cases, we find that for each $M_{\bullet}$ there is an upper limit on the mass of the 
envelope above which the condition $\eta > \eta_{\rm crit}$ is not satisfied, i.e. a 
rotationally-supported accretion flow cannot form.
These limits are shown as thick solid lines with downward pointing arrows in Figure 
\ref{fig_mbh_mstar}. We note that the higher limits correspond to faster surface rotation and 
larger values of $|\beta|$. Fitting the upper limit lines, we obtain that a disc \emph{cannot} 
form for:
\begin{equation}\label{eq_limit}
 M_{\star}\gtrsim 0.9-1.3 \times 10^{5}~{\rm M}_{\sun} \left(\frac{M_{\bullet}}{100~{\rm M}_{\sun}}\right)^{0.82},
\end{equation}
where the $\approx 0.7$ uncertainty factor accounts for differences due to the parameters described 
above. It is very interesting to note that most of the allowed region coincides with 
the Evaporation Strip 
(where the black hole has no time to grow), while the Growth 
Region (where the central black hole could
quickly grow to large masses) is almost entirely excluded.
The sub-linear scaling in equation (\ref{eq_limit}), close to the lines of constant $q$, is the result of the dependence\footnote{As it can be noticed from equation (\ref{eq:diff_rotation_qs}), the shape of the $\Omega$ contours depends mostly on $q$ for different quasi-stars, being $\beta$ constant as physically estimated in Section \ref{subsec_internal_rotation}.
See also Figure \ref{fig_parameters}.} of $\eta$ on $q$, combined with the milder dependence of $\eta_{\rm crit}$ on $q$ and on the properties of the quasi-star.

The models of Figure \ref{fig_mbh_mstar} have been calculated using $b=5$ (see \citetalias{fiacconi+16}).
As we show in the lowest panel of Figure \ref{fig_parameters}, this represents the most favourable case for the formation of a central accretion disc.
For $b = 2$ or 3, the value of $\eta$ crosses $\eta_{\rm crit}$ only for central black holes with mass $\gtrsim 10^4$~M$_{\sun}$, while for $r_{0} = r_{\rm B}$ this does not happen for any mass ratio.
Therefore, in this last case, quasi-stars might not even form in the Evaporation Strip.

Our conclusions might affect the evolution of quasi-stars.
An accretion disc is required as it provides an efficient source of luminosity to sustain the 
envelope through transport of angular momentum and the extraction of gravitational potential 
energy (e.g. via magneto-rotational instability; \citealt{balbus+91}).
If a rotationally-supported disc cannot form, we would be in the presence of an 
optically-thick, quasi-radial flow, that would nearly follow a Bondi-like accretion flow, if well 
within the trapping radius \citep{begelman+78,begelman+79}. In this case, within the Bondi radius, 
the gas becomes supersonic and almost free-falling, converting most of the gravitational potential 
energy into kinetic energy and little into internal energy that could be eventually radiated or 
convectively transported outward.
Since the black hole has no surface, this kinetic energy cannot be dissipated and is advected into 
the black hole. Even assuming a dissipation mechanisms within the flow, most of the 
radiation produced would be dragged inward and swallowed by the black 
hole \citep{begelman+79,alexander+14}.
Therefore, a consistent model for a quasi-star seems not to exist in 
the Growth Region of the parameter space of Figure \ref{fig_mbh_mstar}.

Stepping onto more speculative grounds, we may foreseen 
a possible fate for supermassive black holes that might form in such conditions. 
According to recent numerical and analytical calculations, when a very massive star ($\gtrsim 10^5$~M$_{\sun}$)
forms as consequence of a high accretion rate of gas ($\gtrsim 1$~M$_{\sun}$~yr$^{-1}$), its inner core eventually 
collapses, presumably into a small ($\sim 100$~M$_{\sun}$) black hole \citep{begelman+10,hosokawa+13}.
At this point, however, our results suggest that the surrounding mass might start to be radially accreted, 
unimpeded by the black hole energy feedback. Since there is no maximum limit for the accretion 
rate onto a black hole (there is only a limit in luminosity), this may lead to a phase of 
super-exponential accretion \citep{alexander+14}. The process will stop 
when/if the angular momentum in the accretion flow increases outward so that the circularisation 
radius increases faster than the black hole's $r_{\rm isco}$. The outcome clearly depends on the exact 
hydrodynamics of the flow, but direct formation of massive seeds $\gtrsim 10^5$~M$_{\sun}$ might be 
in principle possible. In this scenario, limiting factors for the black hole seed mass would be 
linked to the galaxy ability to funnel and accumulate pristine gas in its centre: low
cosmological gaseous inflow rate, non efficient angular momentum redistribution and copious star 
formation \citep{latif+13,choi+13,choi+15}.

Small black holes born in the Evaporation Strip might face a different fate.
There, an accretion disc can still form, but the available angular momentum is usually low, such that dissipation 
should occur very close to the black hole. 
One may therefore speculate that a quasi-spherical, geometrically-thick, radiation-dominated 
accretion disc, 
such as a ``ZEBRA'' (ZEro-BeRnoulli Accretion flow; \citealt{coughlin+14}) can form.
Therefore, we have tried to smoothly join the ZEBRA with the envelopes of the models from \citetalias{fiacconi+16} 
at the inner radius $r_{0}$. First, we note that $\eta^2 = \ell_0^2 / \ell^2_{\rm K}(r_0)$
corresponds to the 
normalisation of the specific angular momentum of the gas ``$a$'' (see equation (10) in 
\citealt{coughlin+14}).
Given $\gamma = 4/3$, this only depends on the radial slope $n$ of the mass flow within the ZEBRA 
(i.e. $\dot{M} \propto r^{n}$), which is the main structural parameter of the model. Since 
the ZEBRA should form in the central region of the envelope within $r_{0}$, we assume that its 
external radius $\mathcal{R} = r_{0}$.
Finally, we normalise the density by requiring that the luminosity transported by convection outward 
through the ZEBRA envelope, i.e. 
$L_{\rm adv} \approx 4 \pi \mathcal{R}^2 P(\mathcal{R}) c_{\rm s}(\mathcal{R}) = 4 \pi \mathcal{R}^2 P_0 c_{\rm s,0}$, 
is equal to the central luminosity $L_{\bullet}$ required to self-consistently sustain the envelope.
Unfortunately, we find no consistent solutions where the accretion disc is less 
massive than the black hole, as it is envisaged in the original model.
We would therefore need an extension of this model to self gravitating disc, to assess its viability in our case.
Another possibility is to relax the requirement that $L_{\rm adv} = L_{\bullet}$ and 
speculate instead that $L_{\bullet}$ is provided by partially tapping the energy funnelled into a 
powerful jet, whose presence is foreseen in the ZEBRA model.
However, the jet is likely going to pierce the envelope, behaving as an outlet of energy, and 
therefore how enough energy could be transferred in a gentle, uniform way to the envelope is 
unclear, though possible in principle.

Nonetheless, even if it would be possible to inject within the quasi-star the required 
luminosity at/above the Eddington 
limit for the whole mass, the evaporation of the envelope would anyway prevent substantial accretion 
to occur.
Therefore, there might be two populations of supermassive black hole seeds from direct 
collapse via supermassive stars: one extremely massive, say $>10^{4-5}$~M$_{\sun}$ in massive 
haloes $\gtrsim 10^{8-9}$~M$_{\sun}$, and one extremely light $\sim 100$~M$_{\sun}$ in more 
common haloes at the epoch of formation ($z\sim 15$).
This possibility represents also a ``smooth'' transition between scenarios of light-seed 
formation based on PopIII
stars and massive-seed formation based on direct collapse.


\subsection{Limitations of our treatment} \label{subsec_limitations}

Though intriguing and possible in principle, the speculations discussed in Section \ref{subsec_implications}
relay on results strongly dependent on the assumed model for the quasi-star hydrostatic structure and rotation.
We therefore comment on the limitations of this model.

The simplified description of the quasi-star internal structure as a loaded polytrope (which is unrelated to rotation) requires three parameters to be specified, namely the central pressure $P_{\rm 0}$, the black hole mass $M_{\bullet}$ and the mass of the envelope $M_{\star}$ through the ratio $q$.
However, this neglects the energy production mechanism at the centre, which would introduce an additional relation between e.g. $P_{\rm 0}$ and $M_{\star}$, leaving only two parameters to describe the model with simple scalings (e.g. equations (8)-(10) from \citealt{dotan+11}).
Nonetheless, this treatment provides the correct estimates as long as, for each $M_{\bullet}-q$ 
pair, $P_{\rm 0}$ is chosen consistently with detailed equilibrium models\footnote{Since the self 
consistent
models solve for the energy transport, choosing a consistent value of $P_0$ given $M_{\star}$ and 
$M_{\bullet}$ is then 
implicitly equivalent to consider the  energy transport within the star.
Moreover, the convective envelope of equilibrium models is formally 
obtained by solving the equations of a loaded polytrope, 
and since it dominates the mass and volume of the stars, it provides alone a remarkable description 
of the entire hydrostatic structure.} (e.g. \citetalias{fiacconi+16}).

We model the rotation inside the convective envelope of a quasi-star using the model proposed 
by \citet{balbus+09} and \citet{balbus+10}.
Despite the remarkable agreement with the available data of the internal rotation in the solar 
convective zone and the physical argumentations supporting its reliability, there are no 
\emph{a priori} reasons why this model should apply within a quasi-star nor it should produce a 
sensible description of its rotation, specially at its centre.
However, we can test the fundamental assumption behind it, namely that convective cells are long lived 
compared to the rotation period $t_{\rm rot} \sim 2 \pi (\epsilon \Omega_{\rm K, \star})^{-1}$.
Since convection produces subsonic motions without net mass redistribution, a rough lower limit for
the lifetime of a convective element could be $t_{\rm conv} \sim R_{\star} / c_{\rm s, 0}$.
However, we can obtain a better estimate by applying the mixing length theory \citep[e.g.][]{bohm-vitense+58}, 
which leads to $t_{\rm conv} \sim \sqrt{\alpha~h_{P} / (g \delta)}$, where $\alpha \sim 1-2$ is the
mixing length parameter\footnote{Mixing-lenght theory assumes that convective cells live and mix over a mean free path 
$l = \alpha h_{P}$, where $\alpha$ is a free parameter \citep{bohm-vitense+58}.} 
\citep{asida+00,girardi+00,palmieri+02,ferraro+06}, $h_{P}$ is the pressure scale-height, $g \sim G M_{\star} / R_{\star}^2$ 
is the gravitational field, and $\delta=\Delta T / T$ is the relative (positive) deviation of the temperature gradient from the
adiabatic one in convective regions.
The latter is usually tiny, ranging from $\sim 10^{-5}$ to $\sim 10^{-8}$ in deep convective zones \citep[e.g.][]{bohm-vitense+92,chabrier+07,prialnik+09}, and in fact it justifies the 
description of convective regions through adiabatic relations.
Comparing convection and rotation timescales, we obtain: 
\begin{equation}
\frac{t_{\rm conv}}{t_{\rm rot}} \sim \frac{\epsilon \sqrt{\alpha}}{2 \pi} \left( \frac{h_{P}}{R_{\star} \delta} \right)^{1/2} \sim 
3.6~\epsilon_{0.5}~\delta_{-5}^{-1/2}~(h_{P}/R_{\star})^{1/2}_{-2},
\end{equation}
where $\epsilon = 0.5 \epsilon_{0.5}$, $\alpha = 2$, $\delta = 10^{-5} \delta_{-5}$, and $h_{P}/R_{\star} = 10^{-2} (h_{P}/R_{\star})_{-2}$, as we typically find $h_{P}/R_{\star} \gtrsim 0.01$ in the convective envelope of the models of \citetalias{fiacconi+16}.
This order of magnitude calculation suggests that our model should be reasonably applicable to quasi-stars since the convection timescale is at least comparable or even longer than the rotation period, making convective features long-lived enough to couple with and lie along constant $\Omega$ contours. 

The derivation of equation (\ref{eq_TWE}) formally requires the assumption that rotation is weak, i.e. that departures from sphericity in the hydrostatic equilibrium equation are negligible.
In fact, this assumption enters only in the final substitution $(1/\rho) \partial P / \partial r \rightarrow -{\rm d} \Phi / {\rm d}r$, but it would generally hold in the central regions we are interested in.
Indeed, simple calculations show that the ratio between the centrifugal and the gravitational force becomes smaller toward radii $r \ll R_{\star}$ for any reasonable density profile and angular velocity with a radial scaling shallower than the Keplerian one (see also e.g. \citealt{chandrasekhar+33,monaghan+65}).
Therefore, we conclude that the assumption of weak rotation is not crucial for our findings.

Our model describes a steady-state configuration (thought to be an average in time), whose velocity field is dominated 
by the azimuthal component, i.e. rotation itself.
However, convective regions in differentially rotating zones might lead to features that this approach cannot capture, such as meridional circulation and convective turbulence \citep[e.g.][]{browning+04,ballot+07,browning+08,featherstone+15}.
Those processes are though to be relevant, especially when coupled with magnetic fields, to understand the long-term maintenance and the mutual powering of the differential rotation and the magnetic dynamo within the Sun.
In our case, they might be relevant in the redistribution of angular momentum within the convective envelope, possibly 
having an effect on the rotation of the central region.

Finally, we recall that quasi-stars are thought to be accreting at high rates ($\gtrsim 1$~M$_{\sun}$~yr$^{-1}$) from the local environment.
That means that quasi-stars may not be steady rotating objects, as assumed by our model.
Moreover, accretion from outside may proceed either from a surrounding disc, especially when small scale turbulence is accounted for 
\citep{latif+13}, or in a more disordered fashion from filamentary structures carrying angular momentum with various orientations and amplitudes \citep{choi+15}.
In both cases, gravitational torques from non-axisymmetric features might play a relevant role in influencing the redistribution of angular momentum within the quasi-star envelope, though we cannot explicitly account for that in the present work by assuming a steady-state, temporarily-averaged rotation.


\begin{figure}
\begin{center}
\includegraphics[width=\columnwidth]{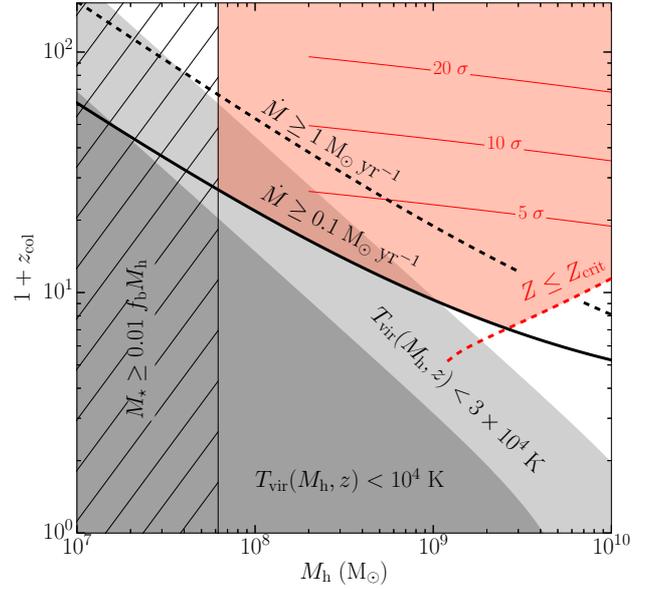}
\end{center}
\caption{Redshift of collapse $z_{\rm col}$ (defined by $F = 0.05$) as a function of the halo mass $M_{\rm h}$.
The continuous and  dashed lines refer to $\dot{M} = 0.1$ and 1~M$_{\sun}$~yr$^{-1}$, respectively.
The dark and light grey regions mark where the halo virial temperature $T_{\rm vir} < 10^4$~K and $T_{\rm vir} < 3 \times 10^4$~K, respectively.
The hatched region marks the halo masses for which a supermassive star $M_{\star} = 10^5$~M$_{\sun}$ represents more than 1\% of the baryonic mass $f_{\rm b} M_{\rm h}$.
The red dashed line marks the redshift threshold above which a halo $M_{\rm h}$ has $Z < Z_{\rm cr}$ after collapsing.
The red shaded region shows where supermassive stars $> M_{\star}$ could form in the $M_{\rm h}-z_{\rm col}$ plane,
and the thin red lines denotes when haloes represent 5, 10, and 20 $\sigma$ over-density fluctuations.}
\label{fig_z_collapse}
\end{figure}

\subsection{Direct collapse haloes}

We now attempt to identify the haloes that could host a supermassive star $M_{\star} \gtrsim 10^{5}$~M$_{\sun}$.
First, we require that $\dot{M} \gtrsim 0.1$~M$_{\sun}$~yr$^{-1}$, as needed to assemble $M_{\star}$ within 
$\sim 1-2$~Myr, i.e. the lifetime of a supermassive star \citep{begelman+10,hosokawa+13}.
The supermassive star may accrete either gas transported through the protogalactic disc or from cosmological 
inflows onto the halo, proceeding all the way down to the centre as cold flows (e.g. \citealt{dimatteo+12}).
The latter case can be translated in a lower limit on the redshift $z$ at which a halo $M_{\rm h}$ can accrete 
at $\dot{M} > 0.1$~M$_{\sun}$~yr$^{-1}$ through the relation\footnote{The usage of equation (\ref{eq_mdot_cosmo}) 
assumes that the gas accretion rate onto the halo is comparable to that onto the forming supermassive star.
However, different assumptions for $\dot{M}(z, M_{\rm h})$ (e.g. Jeans mass collapse over the dynamical time 
$\dot{M} = c_{\rm s}^3/G$ at the virial temperature $T_{\rm vir}$) do not significantly change the result.
Yet, we caution that this approach, in order to keep the calculations simple, neglects the possibility that the 
supermassive star is at the centre of a protogalaxy.} \citep{dekel+09}:
\begin{equation}\label{eq_mdot_cosmo}
\dot{M} \approx 75 \left( \frac{f_{\rm b}}{0.16} \right) \left( \frac{1+z}{3} \right)^{2.25} \left( \frac{M_{\rm h}}{10^{12}~{\rm M}_{\sun}} \right)^{1.15}~{\rm M_{\sun}~yr^{-1}},
\end{equation}
where $f_{\rm b}$ is the cosmic baryon fraction.
We follow \citet{schneider+15} to calculate the collapse redshift of the halo, i.e. the redshift $z_{\rm col}$ at which a fraction $F$ of the mass $M_{\rm h}$ at redshift $\tilde{z}$ is assembled, by solving the following equation for $z_{\rm col}$:
\begin{equation}
\frac{1}{D(z_{\rm col})} = \frac{1}{D(\tilde{z})} + \sqrt{\frac{\pi}{2}} \frac{\sqrt{\sigma^{2}(F M_{\rm h}) - \sigma^{2}(M_{\rm h})}}{\delta_{\rm c, 0}} ,
\end{equation}
where $D(z)$ is the linear growth factor ($D(0) = 1$), $\delta_{\rm c,0} = 1.686$, and $\sigma^{2}(M)$ is the present-day 
variance of the matter density field (i.e. the integral of the linear matter power spectrum over the wavenumber $k$) 
at mass scale $M$ (for additional details, see \citealt{schneider+15}).
Assuming $F = 0.05$, Figure \ref{fig_z_collapse} shows $z_{\rm col}$ as a function of $M_{\rm h}$ for two values of $\dot{M}$.
We adopt the cosmological parameters from the latest Planck results \citep{planck+15} and we find differences within a factor 2 when we vary $F$ from 0.05 to 0.5.

As a second constraint, we require a metallicity below $\log (Z_{\rm cr} / Z_{\sun}) = -3.8$, where $Z_{\sun}$ is the solar metallicity and the critical value roughly corresponds to the transition from PopIII to second population stars \citep{valiante+16}.
We impose this condition by using the stellar mass-metallicity relation as a function of time determined by \citet{savaglio+05}, and then connecting the stellar mass to $M_{\rm h}$ through the halo mass-stellar mass relation from \citet{moster+13}.
After computing the redshift of collapse, we obtain a lower limit $z_{\rm col}(M_{\rm h})$ for haloes $M_{\rm h}$ 
that have $Z < Z_{\rm cr}$ by the end of the collapse.
Finally, the supermassive star cannot be larger than a fraction $f$ of the baryonic mass of the halo, namely $M_{\rm h} > M_{\star} / (f_{\rm b} f)$, where $f \sim 0.01$.
The value of $f$ is chosen in fair agreement with the results of cosmological simulations of the collapse of massive clouds at the centre of dark matter haloes with virial temperature $T_{\rm vir} \gtrsim 10^4$~K (e.g. \citealt{regan+09,latif+13b,choi+15}).

The red shaded region in Figure \ref{fig_z_collapse} shows where all these conditions are satisfied in the $M_{\rm h}-z_{\rm col}$ plane.
We also compare this region with those occupied by haloes with virial temperature $T_{\rm vir} < 10^4$~K and $T_{\rm vir} < 3 \times 10^4$.
The virial temperature is calculated as $T_{\rm vir} \approx (G M_{\rm h} H \sqrt{\Delta/54})^{2/3} \mu m_{\rm p} / k_{\rm B}$,
where $k_{\rm B}$ is the Boltzmann constant, $m_{\rm p}$ is the proton mass, $\mu \approx 0.59$ is the mean molecular weight for ionised hydrogen, $H(z)$ is the Hubble parameter, and $\Delta(z)$ is the $z$-dependent virial over-density \citep{bryan+98}.

The latest haloes that might be able to host a supermassive star $M_{\star} > 10^{5}$~M$_{\sun}$ collapse at $z_{\rm col} \sim 6.5$ and have masses $M_{\rm h} \sim 2-3 \times 10^{9}$~M$_{\sun}$.
Those objects represents $\sim 2\sigma$ peaks in the matter density distribution and have typical comoving number densities $\sim 0.6-0.9$~cMpc$^{-3}$~dex$^{-1}$.
Supermassive stars can also form within both heavier and lighter haloes virtually at any redshift $z_{\rm col} > 10$, when they are able to sustain the inflow and the gas is still pristine enough.
However, beyond $z_{\rm col} \sim 20$, the candidate hosts of supermassive stars more massive than $10^5$~M$_{\sun}$ becomes extremely rare, representing more than $5\sigma$ over-density fluctuations of the matter density field. 
Therefore, we can grossly identify the hosts of supermassive stars possibly leading to the formation of 
$\sim 10^{4-5}$~M$_{\sun}$ back hole seeds as dark mater halos with masses about $\sim 10^{9}$~M$_{\sun}$, collapsing
between $z\sim 20$ and $z\sim 10$, in agreement with previous results (e.g. \citealt{begelman+06,volonteri+10,valiante+16}).
However, we note that our approach (i) requires to extrapolate the used relations to relatively high $z$, and (ii) it does 
not account for environmental effects (e.g. the proximity of a massive halo producing H$_{2}$-dissociating Lyman-Werner photons), therefore the limits above should be taken as approximated.


\subsection{Summary and conclusions}

In this paper, we make a first attempt to discuss possible effects that rotation may have on the structure and evolution
of quasi-stars.
Specifically, we have addressed the issue of whether the redistribution of angular momentum inside the convective envelope of 
a quasi-star in steady rotation may favour the formation of a central accretion disc.
We adopt a model developed initially by \citet{balbus+09} and then improved in a sequence of more recent
papers by the same authors to describe the distribution of angular momentum within the convective zone of the Sun and 
we apply it to quasi-stars.

Within the limitations of this approach (discussed in Section \ref{subsec_limitations}), we find that, at given $M_{\bullet}$, most of the massive quasi-stars might not be able to form a central, rotationally-supported accretion region, while the contrary is true for lower mass quasi-stars, typically living within the Evaporation Strip.
This bimodal behaviour could lead to different fates, depending on the mass of the original supermassive star at the collapse of 
the central core that leads to the formation of the central embryo black hole.
At high masses, the black hole might swallow most of the mass that is still infalling from larger radii without providing enough 
feedback either to stabilise the structure or to halt the collapse.
The central black hole would then accrete a large fraction of the envelop mass, possibly reaching $M_{\bullet} \sim 10^{4-5}$~M$_{\sun}$.
On the other hand, less massive envelopes might be able to form a central accretion disc and to reach an equilibrium 
configuration, i.e. a quasi-star.
However, outflows then suppress the growth of the central black hole, leading to $M_{\bullet} \sim 10^{2-3}$~M$_{\sun}$.

Our results are therefore intriguing, implying possible alternative outcomes for the formation of supermassive black hole 
seeds by direct collapse.
However, this potential needs to be further scrutinised with detailed numerical simulations, as the limitations of our 
analytical treatment suggest caution.
Nonetheless, our first exploration still recommends that further work should be devoted during the future to the topic of rotation 
within supermassive and quasi-stars, since it might be instrumental to better understand crucial details of the formation process
of massive black hole seeds via direct collapse.


\section*{Acknowledgements}

We thank the anonymous Referee for useful comments that helped us improve the quality of this work.
We thank Mitch Begelman, Lucio Mayer and Athena R. Stacy for useful discussions and for a thorough reading of this manuscript in the draft phase.
D.F. is supported by the Swiss National Science Foundation under grant \#No. 200021\_140645.


\bibliographystyle{mnras}
\bibliography{fiacconi_rossi_3}


\bsp	
\label{lastpage}
\end{document}